\documentclass[preprint,12pt]{elsarticle}

 \usepackage{graphicx}
  \graphicspath{ {./} }
 \usepackage{float}

\usepackage{amssymb}
 \usepackage{amsmath}

\usepackage{lineno}
\usepackage{hyperref}
\usepackage{tabularx}
\usepackage{array}

\journal{Journal of Quantitative Spectroscopy and Radiative Transfer}

\begin{document}

\begin{frontmatter}



\title{Improved Recursive Computation of Clebsch-Gordan Coefficients}


\author{Guanglang Xu}
\address{guanglang.xu@helsinki.fi}
\address{Helsinki, Finland}

\begin{abstract}

Fast, accurate, and stable computation of the Clebsch-Gordan (C-G) coefficients is always desirable, for example, in light scattering simulations, the translation of the multipole fields, quantum physics and chemistry. Current recursive methods for computing the C-G coefficients are often unstable for large quantum numbers due to numerical overflow or underflow. In this paper, we present an improved method, the so-called sign-exponent recurrence, for the recursive computation of C-G coefficients. The result shows that the proposed method can significantly improve the stability of the computation without losing its efficiency, producing accurate values for the C-G coefficients even with very large quantum numbers.   
 
\end{abstract}


\end{frontmatter}


\section{Introduction}
\label{S:1}

The Clebsch-Gordan (C-G) coefficients arise whenever the coupling of two angular momenta is involved. The computation of C-G coefficients has a broad range of applications including particle light scattering simulations \cite{mishchenko1991light, wielaard1997improved, mackowski1996calculation, xu1998efficient}, fast multipole methods \cite{danos1965multipole, epton1995multipole}, spherical polar Fourier transform \cite{ritchie2005high}, quantum physics and chemistry \cite{schulten1975semiclassical}. In the context of light scattering by small particles, for instance, the C-G coefficients are needed for the realization of analytical random orientation average using T-matrix. Developed by M. Mishchenko \cite{mishchenko1991light}, the analytical random orientation scheme is perhaps one of the greatest advantages of T-matrix method, which could save massive amount of computational time if the orientation averaged properties are needed. In addition, the C-G coefficient also arises when one need to compute the translation of multipole fields\cite{xu1998efficient}. Consequently, the importance of obtaining fast, reliable and accurate computation of C-G coefficients should not be underestimated. Currently, the computation of C-G coefficients with large quantum numbers is often based on a modified recursion method, which was originally proposed by Schulten and Gordon \cite{schulten1976recursive, schulten1975exact} and later modified and implemented by M. Mishchenko \cite{mishchenko1991light} and Wielaard et al.\cite{wielaard1997improved} in T-matrix simulations. According to \cite{mishchenko2002scattering}, the modified recursive method can compute the C-G coefficients with quantum number up to 150 in a stable and accurate manner. Although the size of T matrix up to this order is already quite large, it is always desirable to develop a highly stable and accurate method for calculating the C-G coefficients with quantum number as large as possible. In this paper, we present an improved method which could significantly extend this limitation without losing the accuracy and efficiency. 

The paper is organised as follows. In section 2, we briefly introduce the widely applied recursion method for the computation of C-G coefficients and analyse its limitations. Section 3 presents our method. In section 4, we demonstrate the stability, accuracy, and efficiency of the proposed method by comparing the results with those from the previous methods. Section 5 concludes this study. 

\section{The recursive computation of C-G coefficients }

The C-G coefficients are related to the Wigner's 3j symbol in accordance with
\begin{equation} \label{eq:cg}
 C\begin{pmatrix}
j_{1} &j_{2}  &j_{3} \\ 
 m_{1}&m_{2}  &-m_{3} 
\end{pmatrix}=(-1)^{j_{1}-j_{2}+m_{3}}\sqrt{2j_{3}+1} \begin{pmatrix}
j_{1} &j_{2}  &j_{3} \\ 
 m_{1}&m_{2}  &m_{3} 
\end{pmatrix},
\end{equation}
where the $j_{1,2,3}$ and $m_{1,2,3}$ are the so-called principle and magnetic angular-momentum quantum numbers respectively. The C-G coefficient will be zero unless the following conditions are satisfied simultaneously, 
\begin{equation}\label{eq:c1}
\left | j_{1}-j_{2} \right |\leq j_{3}\leq (j_{1}+j_{2}), 
\end{equation}
\begin{equation}\label{eq:c2}
m_{3}=-(m_{1}+m_{2}). 
\end{equation}
To be consistent with Schulten and Gordon \cite{schulten1976recursive} on notation, we focus on the discussion of Wigner's $3j$ symbol, while the C-G coefficients can be easily derived from Eq. \ref{eq:cg}.  For small quantum numbers, the $3j$ symbols can be computed conveniently using Racah's formula \cite{racah1942theory}. For large quantum numbers, however, the computation using Racah's formula becomes very expansive and suffers numerical instabilities due to the high order factorials. It is believed that the $3j$ symbols with large quantum numbers can be evaluated most efficiently with the recurrence relations. In light scattering simulations, one needs to compute the $3j$ symbols with a range of principle quantum number $j$ or magnetic quantum number $m$. Because the symmetry properties of the $3j$ symbols (see Appendix D of \cite{mishchenko2002scattering}), the recurrence relations for one of the three quantum numbers shall be enough to compute the recursion of all others. Without loss of generality, let us focus on the computation of $j_{3}$ and $m_{2}$. The recurrence relations for the two quantum numbers are \cite{schulten1976recursive}:
   
\begin{equation}\label{eq:Rcj}
\begin{split}
j_{3}A(j_{3}+1)\begin{pmatrix}
j_{1} &j_{2}  &j_{3}+1 \\ 
m_{1} &m_{2}  &m_{3} 
\end{pmatrix}+B(j_{3})\begin{pmatrix}
 j_{1}& j_{2} & j_{3}\\ 
m_{1} &m_{2}  &m_{3} 
\end{pmatrix}+\\
(j_{3}+1)A(j_{3})\begin{pmatrix}
j_{1} &j_{2}  &j_{3}-1 \\ 
m_{1} &m_{2}  &m_{3} 
\end{pmatrix}=0, 
\end{split}
\end{equation}
\begin{equation}\label{eq:Rcm}
\begin{split}
C(m_{2}+1)\begin{pmatrix}
j_{1} &j_{2}  &j_{3} \\ 
m_{1} &m_{2}+1  &m_{3}-1 
\end{pmatrix}+D(m_{2})\begin{pmatrix}
 j_{1}& j_{2} & j_{3}\\ 
m_{1} &m_{2}  &m_{3} 
\end{pmatrix}+\\
C(m_{2})\begin{pmatrix}
j_{1} &j_{2}  &j_{3} \\ 
m_{1} &m_{2}-1  &m_{3}+1 
\end{pmatrix}=0, 
 \end{split}
\end{equation}
where 
\begin{equation}
A(j_{3})=\left ( [(j_{3})^{2}-(j_{1}-j_{2})^{2}][(j_{1}+j_{2}+1)^{2}-(j_{3})^{2}][(j_{3})^{2}-(m_{3})^{2}] \right )^{1/2},
\end{equation}
\begin{equation}
B(j_{3})=-(2j_{3}+1)[j_{1}(j_{1}+1)m_{3}-j_{2}(j_{2}+1)m_{3}-j_{3}(j_{3}+1)(m_{2}-m_{1})],
 \end{equation}
\begin{equation}
C(m_{2})=[(j_{2}-m_{2}+1)(j_{2}+m_{2})(j_{3}+m_{3}+1)(j_{3}-m_{3})]^{1/2},
\end{equation}
\begin{equation}
D(m_{2})=j_{2}(j_{2}+1)+j_{3}(j_{3}+1)-j_{1}(j_{1}+1)+2m_{2}m_{3}.
\end{equation}
For recursive computation using Eq.(4), $j_{3}$ lies in the following range
\begin{equation}
\max(|m_{1}+m_{2}|,|j_{1}-j_{2}|)\leq j_{3}\leq (j_{1}+j_{2}).
\end{equation}
For recursive computation using Eq.(5), $m_{2}$ lies in the following range  
\begin{equation}
-\min(j_{2}, j_{3}+m_{1})\leq m_{2}\leq \min(j_{2}, j_{3}-m_{1}).
\end{equation}
In the original method proposed by Schulten and Gordon \cite{schulten1976recursive}, the recursions can start from arbitrary real number, e.g., $unity$, and go both forward and backward from the minimum and maximum quantum number respectively. The method therefore requires the computation of a scaling factor such that the forward and backward recursions give the same number at the intermediate quantum number. The coefficients can then be determined by applying the unitary properties: 
\begin{equation}
\sum _{j_{3}=j_{3_{\min}}}^{j_{3_{\max}}} (2j_{3}+1)\begin{pmatrix}
j_{1} &j_{2}  &j_{3} \\ 
m_{1} &m_{2}  &m_{3} 
\end{pmatrix}^{2}=1\\
\end{equation}
\begin{equation}
\sum _{m_{2}=m_{2_{\min}}}^{m_{2_{\max}}}(2j_{1}+1)\begin{pmatrix}
j_{1} &j_{2}  &j_{3} \\ 
m_{1} &m_{2}  &m_{3} 
\end{pmatrix}^{2}=1 
\end{equation}

\begin{figure}[H]
    \centering
    \includegraphics[width=0.85\textwidth]{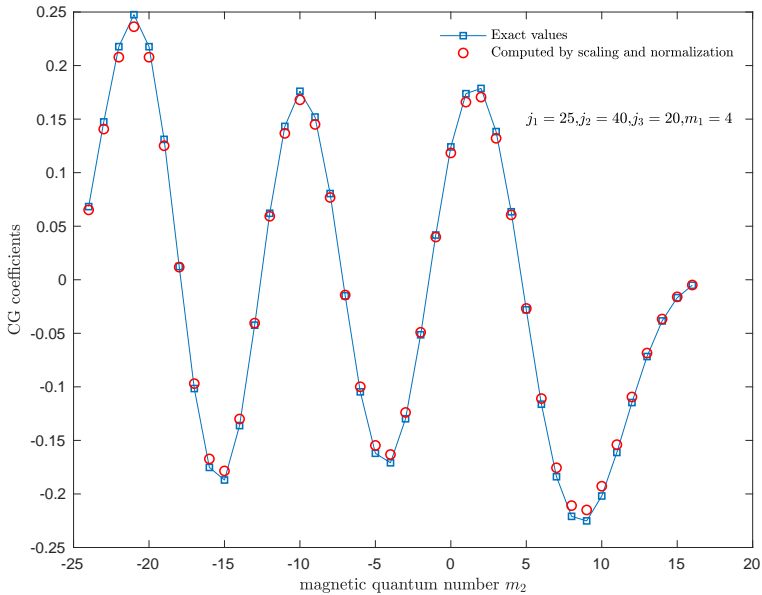}
    \caption{Global shifting caused by scaling and normalization.}
    \label{fig:scaling}
\end{figure}

In addition to iteration direction \cite{schulten1976recursive}, there are two more sources that could cause inaccuracy or numerical instability via recursion. Firstly, because the recursion starts with arbitrary number, the computed values need to be scaled twice, first by the scaling factor and then by the normalization factor. The errors of the factors, possibly arising from the inaccuracy of particular coefficients or their ratios, could propagate to the whole group of the computed values. To illustrate this, we compare the exact values with those computed by scaling and normalization as displayed in Fig. \ref{fig:scaling}. It can be seen that most of the computed values can be shifted by certain magnitude due to the multiplication of the factors.  

The second source of errors in the existing methods is the lack of a mechanism to avoid numerical outflows or underflows. Note that in the case of high order quantum numbers, the magnitude of quantities (or their ratios) involved in the computations could be extremely small or large. 

In \cite{luscombe1998simplified}, Luscombe and Luban propose to iterate the ratio of two successive $3j$ symbols to avoid numerical overflows. Nevertheless, their method is not without drawbacks.  Firstly, it has to perform normailzation, which could cause unnecessary shifting of values. Secondly,  in their method, the values of C-G coefficients are to be obtained by multiplication of many ratios, and this could still induce numerical overflow/underflow,  even though the iteration of the ratios may have no such risks. Thirdly,  the necessity of identifying classical and nonclassical regions complicates the algorithm.

One way to remove the necessity of scaling and normalization is to start with an exact value of C-G coefficient at the minimum or maximum quantum number. As described in \ref{sr2}, there are four different cases to be considered for $j_{3}=j_{3_{\min}}$, while there is only one case for $j_{3}=j_{3_{\max}}$. The usage of exact starting values exclude the necessity of scaling and normalization, which improves the computational stability and accuracy. However, it does not exclude the possibility of numerical underflow/overflow, because for large quantum numbers, the factorial computations for the starting values will likely exceed the precision of the arithmetic. 
\section{The sign-exponent recurrence method}

If the quantum numbers satisfy the conditions of Eq. \ref{eq:c1} and Eq. \ref{eq:c2}, the $3j$ symbols are generally non-zero. However, it is well-known that  
some of the coefficients can be ``accidentally''  zero even if the conditions are fulfilled \cite{heim2009some}. Such zeros are called non-trivial zeros. But the non-trivial zeros are quite rarely encountered, for the moment, let us assume that all the coefficients involved are non-zero. In this case, one can write arbitrary $3j$ symbols as

\begin{equation}\label{eq:exf1}
f(j_{3})=\begin{pmatrix}
j_{1} &j_{2}  &j_{3} \\ 
 m_{1}&m_{2}  &m_{3} 
\end{pmatrix}=s(j_{3})\exp(k(j_{3})),
\end{equation}
\begin{equation}\label{eq:exf2}
g(m_{2})=\begin{pmatrix}
j_{1} &j_{2}  &j_{3} \\ 
 m_{1}&m_{2}  & -m_{2}-m_{1} 
\end{pmatrix}=s(m_{2})\exp(h(m_{2})),
\end{equation}
where $k(j_{3})$ and $h(m_{2})$ are real functions of $j_{3}$ and $m_{3}$ respectively, and 

\begin{equation}\label{eq:rec5}
\begin{cases}
&s(j_{3})=sign(f(j_{3}))\\
&s(m_{2})=sign(g(m_{2}))\\
\end{cases}
\end{equation}

The $sign$ function is defined as 
\begin{equation}\label{eq:sf}
sign(\alpha)= \begin{cases}
+1 & \text{if $\alpha > 0$},\\
0 & \text{if $\alpha = 0$},\\
-1 & \text{if $\alpha < 0$}.
\end{cases}
\end{equation}
By introducing Eq. \ref{eq:exf1} into Eq. \ref{eq:Rcj} , we obtain 
\begin{equation}\label{eq:rec1}
\begin{split}
s(j_{3}+1)\exp(k(j_{3}+1))=\exp(k(j_{3}))[\alpha(j_{3})s(j_{3})\\
+\beta(j_{3})s(j_{3}-1)\exp(-\Delta(j_{3}))], 
\end{split}
\end{equation}
where 
\begin{equation}\label{eq:rec2}
\Delta(j_{3})=k(j_{3})-k(j_{3}-1),
\end{equation}
\begin{equation}\label{eq:rec3}
\alpha(j_{3})=-\frac{B(j_{3})}{j_{3}A(j_{3}+1)},
\end{equation}
\begin{equation}\label{eq:rec4}
\beta(j_{3})=-\frac{(j_{3}+1)A(j_{3})}{j_{3}A(j_{3}+1)}.
\end{equation}

From Eq.\ref{eq:rec1}, one can obtain the recurrence relations for both $s(j_{3})$ and $k(j_{3})$, i.e., 
\begin{equation}\label{eq:rec6}
\begin{cases}
&s(j_{3}+1)=sign[\alpha(j_{3})s(j_{3})
+\beta(j_{3})s(j_{3}-1)\exp(-\Delta(j_{3}))]\\
&k(j_{3}+1)=k(j_{3})+\ln|\alpha(j_{3})s(j_{3})
+\beta(j_{3})s(j_{3}-1)\exp(-\Delta(j_{3}))|. 
\end{cases}
\end{equation}
Similarly,  recurrence relation for $m_{2}$ can be obtained by introducing Eq. \ref{eq:exf2} into Eq. \ref{eq:Rcm}, i.e., 
\begin{equation}\label{eq:rec8}
\begin{cases}
&s(m_{2}+1)=sign[\lambda(m_{2})s(m_{2})
+\eta(m_{2})s(m_{2}-1)\exp(-\Delta(m_{2}))]\\
&h(m_{2}+1)=h(m_{2})+\ln|\lambda(m_{2})s(m_{2})
+\eta(m_{2})s(m_{2}-1)\exp(-\Delta(m_{2}))|\\
\end{cases}
\end{equation}
where 
\begin{equation}\label{eq:rec12}
\Delta(m_{2})=h(m_{2})-h(m_{2}-1),
\end{equation}
\begin{equation}\label{eq:rec13}
\lambda (m_{2}) =-\frac{D(m_{2})}{C(m_{2}+1)},
\end{equation}
\begin{equation}\label{eq:rec14}
\eta (m_{2}) =-\frac{C(m_{2})}{C(m_{2}+1)}.
\end{equation}
At the starting minimum quantum number, we have 
\begin{equation}\label{eq:rec15ad}
\begin{cases}
&s(j_{3_{min}}-1)=0\\
&s(m_{2_{min}}-1)=0.\\
\end{cases}
\end{equation}
Therefore the iteration becomes 
\begin{equation}\label{eq:rec6ad}
\begin{cases}
&s(j_{3_{min}}+1)=sign[\alpha(j_{3_{min}})s(j_{3_{min}})]\\
&k(j_{3_{min}}+1)=k(j_{3_{min}})+\ln|\alpha(j_{3_{min}})s(j_{3_{min}})|,\\ 
\end{cases}
\end{equation}
and
\begin{equation}\label{eq:rec8ad}
\begin{cases}
&s(m_{2_{min}}+1)=sign[\lambda(m_{2_{min}})s(m_{2_{min}})]\\
&h(m_{2_{min}}+1)=h(m_{2_{min}})+\ln|\lambda(m_{2_{min}})s(m_{2_{min}})|.\\
\end{cases}
\end{equation}
It can be seen that we have separated the original recurrence relation into sign-recursion and exponent-recursion. Note that the above relations are valid for the forward direction which increases the quantum number. Similarly, it would be straightforward to derive the backward recurrence relation that reduces the quantum number. The basic idea behind such recurrence relations is that we could focus on computing the sign and exponent of the coefficient, rather than the coefficient itself. In principle, as long as the sign and exponent are computed accurately, the coefficient can always be calculated accurately via Eq. \ref{eq:exf1} or Eq. \ref{eq:exf2}. Because the $3j$ symbols can vary many orders of magnitude, the sign-exponent recurrence can significantly reduce the risk of numerical underflows/overflows. To apply the derived relations, we just need to compute the starting exponent and sign of the coefficient at the minimum or maximum quantum numbers. According to our numerical tests, the method would not induce numerical underflow/overflow even with quantum number larger than $10$ million.

For now, we shall consider the problem of encountering non-trivial zeros. Once zero-value is encountered, the $sign$ value will become $0$ and the exponent will become $-\infty$, making the computation of $\Delta(m_{2})$ or $\Delta(j_{3})$ meaningless. Therefore, the above recurrence relations must be avoided and the original three-term linear recurrence relations should be applied. The condition for using the sign-exponent recursions shall be that both $s(j_{3}-1)$ and $s(j_{3})$ are non-zero (same for $m_{2}$ case), otherwise the  three-term linear relations should be invoked. For large quantum numbers, the non-trivial zeros are rarely encountered. For most of the cases, only the sign-exponent iteration is invoked in the computation. 

\section{Results and Discussion}
\label{S:2}

\begin{figure}[H]
    \centering
    \includegraphics[width=0.96\textwidth]{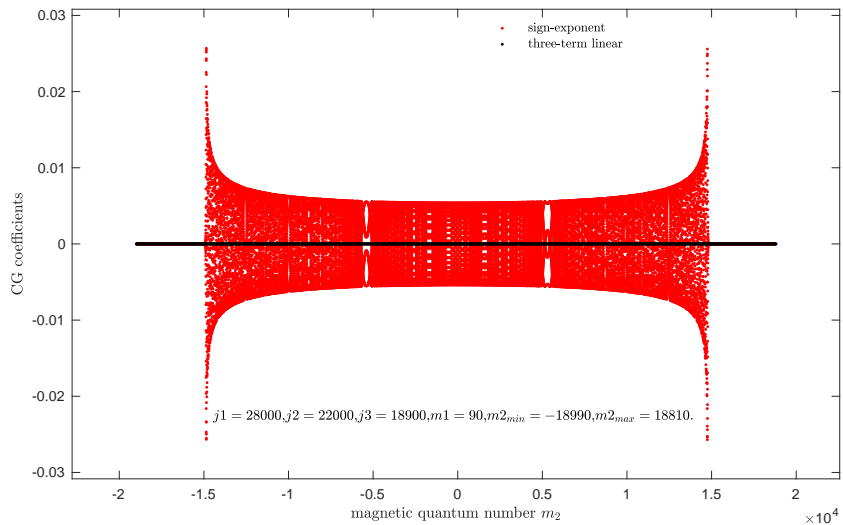}
    \caption{Test comparison between the three-term linear recurrence  and sign-exponent recurrence for large quantum numbers. The values have been scaled back to C-G coefficients.The $red$ dot marks the values from the sign-exponent recurrence method, while the $black$ dot marks the values from the three-term linear recurrence method.}
    \label{fig:sta}
\end{figure}

   In this section, we shall discuss the stability, accuracy, and efficiency of the proposed method by comparing with the widely applied three-term linear recurrence with exact starting values. For large quantum numbers, in general, all recursion-based methods are much faster than those using the direct definition or formula. The efficiency of the proposed method is almost the same as the previous recursion methods. To quantify the accuracy of the computation, we define the following error term, which is consistent with \cite{xu1998efficient}.

\begin{equation}\label{eq:rec15}
R=|1- \sum _{m_{2}=m_{2_{min}}}^{m_{2_{max}}}(2j_{1}+1)\begin{pmatrix}
j_{1} &j_{2}  &j_{3} \\ 
m_{1} &m_{2}  &m_{3} 
\end{pmatrix}^{2}| 
\end{equation}
Please note that all the results are obtained by applying the double-precision arithmetic. We compute the starting values for $m_{2}$ recurrence by firstly using the recurrence relation for $j_{3}$, therefore the comparison on $m_{2}$ recurrence is preferred. The following figure displays a comparison between the three-term linear recurrence and sign-exponent recurrence method. 
\begin{figure}[H]
    \centering
    \includegraphics[width=0.9\textwidth]{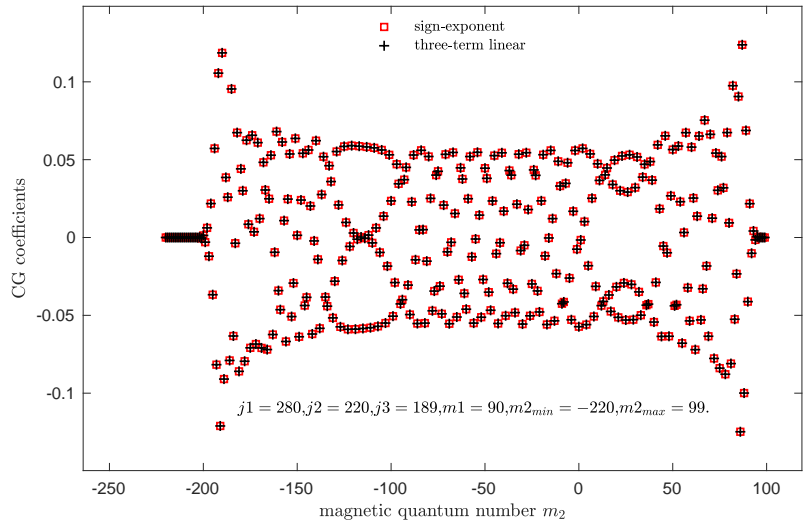}
    \caption{A test comparison between the widely-applied three-term linear recursion and the sign-exponent recursion method introduced in this study. The values are scaled back to C-G coefficients.The black $plus-sign$ marks the values computed by the three-term linear recurrence method, while the red $circle$ marks the values computed by the sign-exponent recurrence method.}
    \label{fig:svt}
\end{figure}
It can be seen that the results of the two are indistinguishable from the figure. In fact, for the three-term linear recursion, $R=1.0334 \times 10^{-12}$, while for the sign-exponent recursion, $R=1.0214 \times 10^{-12}$. The values of $R$ suggest that the sign-exponent recurrence method has at least the same performance as the three-term linear recurrence on numerical accuracy. The biggest advantage of the proposed method perhaps is that it can avoid numerical underflow when computing the starting values and this is crucial for dealing with large quantum numbers because the starting values could be extremely small.  Figure \ref{fig:sta} demonstrates the comparison for the case of very large quantum numbers. The three-term linear recurrence suffers numerical underflow and becomes zero throughout the iteration. On the contrary, the sign-exponent recurrence method excludes the risk of obtaining zero starting values. Consequently,  for the three-term linear recurrence, $R=1$, while for the sign-exponent recursion, $R=5.9769 \times 10^{-10}$, which still maintains high accuracy. Based on our extensive tests, the proposed method is generally much more stable than the original linear recursion method, while having the same level of numerical accuracy and efficiency. To further demonstrate the accuracy and stability of our method, in \ref{sr1}, we provide a link to our Matlab code  and more test examples in comparisons with the most accurate package Python SymPy.    

\section{Conclusion}

In this paper, an improved recursion method for computing the C-G coefficients is introduced. Specifically, the method separates the recursion process into sign-recursion and exponent-recursion, while the C-G value itself is not involved in the recursion except that a non-trivial zero occurs. The C-G values can be obtained after the computation of their signs and exponents. By using the sign-exponent recursion, the method removes the risk of generating numerical overflows or underflows. The results presented in this paper, together with our extensive tests, have shown that the sign-exponent recurrence method is in general more stable than the original three-term linear recurrence method, while having the same level of accuracy and efficiency.            


\appendix

\section{The computation of starting values}
 \label{sr2}
The recursive computation without using the scaling and normalization relies on the exact computation of the starting values. For backward recursion, the starting value is unique, i.e., 
\begin{equation}\label{eq:rec15a}
\begin{split}
&\begin{pmatrix}
j_{1} &j_{2}  &j_{1}+j_{2} \\ 
m_{1} &m_{2}  &-m_{1}-m_{2} 
\end{pmatrix}=(-1)^{j_{1}+j_{2}+m_{2}+m_{1}}\times\\
&\left [\frac{(2j_{1})!(2j_{2})!(j_{1}+j_{2}-m_{1}-m_{2})!(j_{1}+j_{2}+m_{1}+m_{2})!}{(2j_{1}+2j_{2}+1)!(j_{1}-m_{1})!(j_{1}+m_{1})!(j_{2}-m_{2})!(j_{2}+m_{2})!} .  \right ]^{1/2}\\
\end{split}
\end{equation}
For the forward iteration, we have four possibilities, depending on the values of $j_{{3}_{\min}}$, i.e., 

\begin{equation}\label{eq:rec16}
\begin{split}
&\begin{pmatrix}
j_{1} &j_{2}  &j_{1}-j_{2} \\ 
m_{1} &m_{2}  &-m_{1}-m_{2} 
\end{pmatrix}=(-1)^{j_{1}+m_{1}}\times\\
&\left [\frac{(2j_{1}-2j_{2})!(2j_{2})!(j_{1}-m_{1})!(j_{1}+m_{1})!}{(j_{1}-j_{2}-m_{1}-m_{2})!(j_{2}-m_{2})!(j_{2}+m_{2})!(j_{1}-j_{2}+m_{1}+m_{2})!(2j_{1}+1)!}\right ]^{1/2}\\
\end{split}
\end{equation}

\begin{equation}\label{eq:rec17}
\begin{split}
&\begin{pmatrix}
j_{1} &j_{2}  &j_{2}-j_{1} \\ 
m_{1} &m_{2}  &-m_{1}-m_{2} 
\end{pmatrix}=(-1)^{j_{2}+m_{2}}\times\\
&\left [\frac{(2j_{2}-2j_{1})!(2j_{1})!(j_{2}-m_{1})!(j_{2}+m_{1})!}{(j_{2}-j_{1}-m_{1}-m_{2})!(j_{1}-m_{1})!(j_{1}+m_{1})!(j_{2}-j_{1}+m_{1}+m_{2})!(2j_{2}+1)!}\right ]^{1/2}\\
\end{split}
\end{equation}

\begin{equation}\label{eq:rec18}
\begin{split}
&\begin{pmatrix}
j_{1} &j_{2}  &m_{2}+m_{1} \\ 
m_{1} &m_{2}  &-m_{1}-m_{2} 
\end{pmatrix}=(-1)^{j_{2}+m_{2}}\times\\
&\left [\frac{(j_{1}+m_{1})!(j_{2}+m_{2})!(j_{2}+j_{1}-m_{1}-m_{2})!(2m_{1}+2m_{2})!}{(j_{1}-m_{1})!(j_{2}-m_{2})!(j_{1}-j_{2}+m_{1}+m_{2})!(j_{2}-j_{1}+m_{1}+m_{2})!(j_{1}+j_{2}+m_{1}+m_{2}+1)!}\right ]^{1/2}\\
\end{split} 
 \end{equation}

\begin{equation}\label{eq:rec19}
\begin{split}
&\begin{pmatrix}
j_{1} &j_{2}  &-m_{2}-m_{1} \\ 
m_{1} &m_{2}  &-m_{1}-m_{2} 
\end{pmatrix}=(-1)^{j_{1}+m_{1}}\times\\
&\left [\frac{(j_{1}-m_{1})!(j_{2}-m_{2})!(j_{2}+j_{1}+m_{1}+m_{2})!(-2m_{1}-2m_{2})!}{(j_{1}+m_{1})!(j_{2}+m_{2})!(j_{1}-j_{2}-m_{1}-m_{2})!(j_{2}-j_{1}-m_{1}-m_{2})!(j_{1}+j_{2}-m_{1}-m_{2}+1)!}\right ]^{1/2}\\
\end{split} 
 \end{equation}

In the computation of the above values, we can compute the logarithm of the factorials to avoid overflow. To be more specific, one can write arbitrary starting value as
\begin{equation}\label{eq:rec20}
f(j_{start})=s(j_{start})\exp(k(j_{start}))=(-1)^{l}\left [  \frac{a_{1}!a_{2}!a_{3}!a_{4}!}{b_{1}!b_{2}!b_{3}!b_{4}!b_{5}!}\right ]^{1/2}
\end{equation}
Obviously, the starting values for $s$ and $k$ functions are 
\begin{equation}\label{eq:rec21}
s(j_{start})=(-1)^{l}\\
\end{equation}
\begin{equation}\label{eq:rec22}
\begin{split}
&k(j_{start})=\frac{1}{2}  [ \ln(a_{1}!)+\ln(a_{2}!)+\ln(a_{3}!)+\ln(a_{4}!)\\
&-\ln(b_{1}!)-\ln(b_{2}!)-\ln(b_{3}!)-\ln(b_{4}!)-\ln(b_{5}!)) ] 
\end{split} 
\end{equation}
To compute the logarithm of the factorials, we may use the following formula to avoid numerical overflow, 
\begin{equation}\label{eq:rec23}
\ln(N!)=\sum_{i=1}^{N}\ln(i).
\end{equation}

  \section{Comparison with Python library SymPy}
   \label{sr1}

  To demonstrate the accuracy of the proposed method, we compile some of the results from Python library SymPy and compare with the proposed method in this study. The code for our method can be accessed from \cite{Xu:2020}. The code for SymPy can be obtained from \cite{Python:2020}.   
The SymPy is based on symbolic manipulation, which could be considered as the most accurate method. The computational time by the recursion methods ranges between $0.01$ to $0.03$ seconds for most of the cases with quantum number smaller than $1000$, while the SymPy package could take up to roughly a few seconds.  It can be seen that the two recursion methods are both very accurate comparing to SymPy and their levels of accuracy are pretty much the same.  However, as we increase the quantum numbers, the three-term linear recursion becomes unstable and generates zeros due to numerical underflows. The Python SymPy simply produces errors at very large quantum numbers, while the sign-exponent method is still stable and produces reasonable results. \\

 \begin{tabular}{ |p{0.9cm}||p{3.6cm}|p{3.6cm}|p{3.8cm}|  }
 \hline
 \multicolumn{4}{|c|}{$j_{1}=280$, $j_{2}=220$, $j_{3}=189$, $m_{1}=90$} \\
 \hline
 $m_{2}$  & Sign-Exponent  & Three-Term Linear  & Python SymPy \\
 \hline
 $-120$  &  0.002887948213256     &0.002887948213256 &  0.00288794821325701 \\
 $-125$ &   0.058508415288557  & 0.058508415288558   & 0.0585084152885739 \\
 $-128$  & 0.020928845109162 &  0.020928845109162  &  0.020928845109168 \\
 $-130$ & -0.028070293415027 & -0.028070293415028 & -0.0280702934150357\\ 
 $-135$ &-0.038257492867934 & -0.038257492867934 &  -0.038257492867945\\
 \hline
\end{tabular}

 \begin{tabular}{ |p{0.9cm}||p{3.6cm}|p{3.6cm}|p{3.8cm}|  }
 \hline
 \multicolumn{4}{|c|}{$j_{1}=480$, $j_{2}=320$, $j_{3}=300$, $m_{1}=90$} \\
 \hline
 $m_{2}$  & Sign-Exponent  & Three-Term Linear  & Python SymPy \\
 \hline
 $-125$ &   -0.046262518791471  & -0.046262518791468   & -0.0462625187915161\\
 $-128$  & -0.041062246328595 &  -0.041062246328592 &  -0.0410622463286351 \\
 $-120$ & 0.041718263408549 & 0.041718263408546 & 0.0417182634085898\\ 
 $-130$ & 0.047752989423874 & 0.047752989423870 & 0.0477529894239198\\
  $-135$  &  -0.047799931849649    &-0.047799931849645 & -0.0477999318496948 \\
 \hline
\end{tabular}

 \begin{tabular}{ |p{0.9cm}||p{3.6cm}|p{3.6cm}|p{3.8cm}|  }
 \hline
 \multicolumn{4}{|c|}{$j_{1}=700$, $j_{2}=620$, $j_{3}=230$, $m_{1}=300$} \\
 \hline
 $m_{2}$  & Sign-Exponent  & Three-Term Linear  & Python SymPy \\
 \hline
 $-200$ &   -0.029578200668109  & -0.029578200668109   &-0.0295782006677839\\
 $-250$  & -0.033722189882640 & -0.033722189882640 & -0.0337221898822695 \\
 $-300$ & -0.000885723206928 & -0.000885723206928 & -0.0008857232069200\\ 
 $-350$ & 0.032668945677003 & 0.032668945677002 & 0.0326689456767071\\
  $-400$  &  0.032449523658905    &0.032449523658904  & 0.0324495236586107 \\
 \hline
\end{tabular}

 \begin{tabular}{ |p{1.9cm}||p{3.6cm}|p{3.6cm}|p{2.8cm}|  }
 \hline
 \multicolumn{4}{|c|}{$j_{1}=7000$, $j_{2}=6200$, $j_{3}=2300$, $m_{1}=3000$} \\
 \hline
 $m_{2}$  & Sign-Exponent  & Three-Term Linear  & Python SymPy \\
 \hline
 $-2000$ &  0.001244977301861 & 0.001244977301860   & Error. \\
 $-2500$  & -0.007275107384171 & -0.007275107384171 &Error.  \\
 $-3000$ & 0.002712153629703 & -0.000000000000000 & Error.  \\ 
 $-3500$ & 0.006665616564930 & -0.000000000000000 & Error. \\
  $-4000$  & -0.010583441967577   &-0.000000000000000  & Error. \\
 \hline
\end{tabular}

 \begin{tabular}{ |p{1.9cm}||p{3.6cm}|p{3.6cm}|p{2.8cm}|  }
 \hline
 \multicolumn{4}{|c|}{ $j_{1}=9000000$, $j_{2}=620000$, $j_{3}=7800000$, $m_{1}=3000000$} \\
 \hline
 $m_{2}$  & Sign-Exponent  & Three-Term Linear  & Python SymPy \\
 \hline
 $-2000000$ &  -0.000258969176674 & 0.000000000000000  & Error. \\
 $-2500000$  & -0.000236627240189 & -0.000000000000000 &Error.  \\
 $-3000000$ & -0.000146400518567 & -0.000000000000000 & Error.  \\ 
 $-3500000$ & -0.000144722745846 & 0.000000000000000 & Error. \\
  $+3500000$  & 0.000396746931467   &-0.000000000000000  & Error. \\
 \hline
\end{tabular}




\bibliographystyle{model1-num-names}
\bibliography{CG.bib}

\begin{thebibliography}{16}
\expandafter\ifx\csname natexlab\endcsname\relax\def\natexlab#1{#1}\fi
\providecommand{\bibinfo}[2]{#2}
\ifx\xfnm\relax \def\xfnm[#1]{\unskip,\space#1}\fi
\bibitem[{Mishchenko(1991)}]{mishchenko1991light}
\bibinfo{author}{M.~Mishchenko},
\newblock \bibinfo{title}{Light scattering by randomly oriented axially
  symmetric particles},
\newblock \bibinfo{journal}{JOSA A} \bibinfo{volume}{8} (\bibinfo{year}{1991})
  \bibinfo{pages}{871--882}.
\bibitem[{Wielaard et~al.(1997)Wielaard, Mishchenko, Macke, and
  Carlson}]{wielaard1997improved}
\bibinfo{author}{D.~J. Wielaard}, \bibinfo{author}{M.~I. Mishchenko},
  \bibinfo{author}{A.~Macke}, \bibinfo{author}{B.~E. Carlson},
\newblock \bibinfo{title}{Improved t-matrix computations for large,
  nonabsorbing and weakly absorbing nonspherical particles and comparison with
  geometrical-optics approximation},
\newblock \bibinfo{journal}{Applied optics} \bibinfo{volume}{36}
  (\bibinfo{year}{1997}) \bibinfo{pages}{4305--4313}.
\bibitem[{Mackowski and Mishchenko(1996)}]{mackowski1996calculation}
\bibinfo{author}{D.~W. Mackowski}, \bibinfo{author}{M.~I. Mishchenko},
\newblock \bibinfo{title}{Calculation of the t matrix and the scattering matrix
  for ensembles of spheres},
\newblock \bibinfo{journal}{JOSA A} \bibinfo{volume}{13} (\bibinfo{year}{1996})
  \bibinfo{pages}{2266--2278}.
\bibitem[{Xu(1998)}]{xu1998efficient}
\bibinfo{author}{Y.-L. Xu},
\newblock \bibinfo{title}{Efficient evaluation of vector translation
  coefficients in multiparticle light-scattering theories},
\newblock \bibinfo{journal}{Journal of Computational Physics}
  \bibinfo{volume}{139} (\bibinfo{year}{1998}) \bibinfo{pages}{137--165}.
\bibitem[{Danos and Maximon(1965)}]{danos1965multipole}
\bibinfo{author}{M.~Danos}, \bibinfo{author}{L.~Maximon},
\newblock \bibinfo{title}{Multipole matrix elements of the translation
  operator},
\newblock \bibinfo{journal}{Journal of Mathematical Physics}
  \bibinfo{volume}{6} (\bibinfo{year}{1965}) \bibinfo{pages}{766--778}.
\bibitem[{Epton and Dembart(1995)}]{epton1995multipole}
\bibinfo{author}{M.~A. Epton}, \bibinfo{author}{B.~Dembart},
\newblock \bibinfo{title}{Multipole translation theory for the
  three-dimensional laplace and helmholtz equations},
\newblock \bibinfo{journal}{SIAM Journal on Scientific Computing}
  \bibinfo{volume}{16} (\bibinfo{year}{1995}) \bibinfo{pages}{865--897}.
\bibitem[{Ritchie(2005)}]{ritchie2005high}
\bibinfo{author}{D.~W. Ritchie},
\newblock \bibinfo{title}{High-order analytic translation matrix elements for
  real-space six-dimensional polar fourier correlations},
\newblock \bibinfo{journal}{Journal of applied crystallography}
  \bibinfo{volume}{38} (\bibinfo{year}{2005}) \bibinfo{pages}{808--818}.
\bibitem[{Schulten and Gordon(1975)}]{schulten1975semiclassical}
\bibinfo{author}{K.~Schulten}, \bibinfo{author}{R.~G. Gordon},
\newblock \bibinfo{title}{Semiclassical approximations to 3 j-and 6
  j-coefficients for quantum-mechanical coupling of angular momenta},
\newblock \bibinfo{journal}{Journal of Mathematical Physics}
  \bibinfo{volume}{16} (\bibinfo{year}{1975}) \bibinfo{pages}{1971--1988}.
\bibitem[{Schulten and Gordon(1976)}]{schulten1976recursive}
\bibinfo{author}{K.~Schulten}, \bibinfo{author}{R.~Gordon},
\newblock \bibinfo{title}{Recursive evaluation of 3j and 6j coefficients},
\newblock \bibinfo{journal}{Computer Physics Communications}
  \bibinfo{volume}{11} (\bibinfo{year}{1976}) \bibinfo{pages}{269--278}.
\bibitem[{Schulten and Gordon(1975)}]{schulten1975exact}
\bibinfo{author}{K.~Schulten}, \bibinfo{author}{R.~G. Gordon},
\newblock \bibinfo{title}{Exact recursive evaluation of 3 j-and 6
  j-coefficients for quantum-mechanical coupling of angular momenta},
\newblock \bibinfo{journal}{Journal of Mathematical Physics}
  \bibinfo{volume}{16} (\bibinfo{year}{1975}) \bibinfo{pages}{1961--1970}.
\bibitem[{Mishchenko et~al.(2002)Mishchenko, Travis, and
  Lacis}]{mishchenko2002scattering}
\bibinfo{author}{M.~I. Mishchenko}, \bibinfo{author}{L.~D. Travis},
  \bibinfo{author}{A.~A. Lacis}, \bibinfo{title}{Scattering, absorption, and
  emission of light by small particles}, \bibinfo{publisher}{Cambridge
  university press}, \bibinfo{year}{2002}.
\bibitem[{Racah(1942)}]{racah1942theory}
\bibinfo{author}{G.~Racah},
\newblock \bibinfo{title}{Theory of complex spectra. ii},
\newblock \bibinfo{journal}{Physical Review} \bibinfo{volume}{62}
  (\bibinfo{year}{1942}) \bibinfo{pages}{438}.
\bibitem[{Luscombe and Luban(1998)}]{luscombe1998simplified}
\bibinfo{author}{J.~H. Luscombe}, \bibinfo{author}{M.~Luban},
\newblock \bibinfo{title}{Simplified recursive algorithm for wigner 3 j and 6 j
  symbols},
\newblock \bibinfo{journal}{Physical Review E} \bibinfo{volume}{57}
  (\bibinfo{year}{1998}) \bibinfo{pages}{7274}.
\bibitem[{Heim et~al.(2009)Heim, Hinze, and Rau}]{heim2009some}
\bibinfo{author}{T.~Heim}, \bibinfo{author}{J.~Hinze},
  \bibinfo{author}{A.~Rau},
\newblock \bibinfo{title}{Some classes of ‘nontrivial zeroes’ of angular
  momentum addition coefficients},
\newblock \bibinfo{journal}{Journal of Physics A: Mathematical and Theoretical}
  \bibinfo{volume}{42} (\bibinfo{year}{2009}) \bibinfo{pages}{175203}.
\bibitem[{Xu(2020)}]{Xu:2020}
\bibinfo{author}{G.~Xu}, \bibinfo{title}{A Matlab code for the sign-exponent
  recursive computation of CG-coefficients}, \bibinfo{year}{June 2020}.
  \bibinfo{note}{\url{https://github.com/GXhelsinki/Clebsch-Gordan-Coefficients-}}.
\bibitem[{SymPy(2020)}]{Python:2020}
\bibinfo{author}{SymPy}, \bibinfo{title}{Python library SymPy for Computation
  of Clebsch-Gordan Coefficients}, \bibinfo{year}{June 2020}.
  \bibinfo{note}{\url{https://docs.sympy.org/latest/modules/physics/quantum/cg.html}}.

\end{thebibliography}







\end{document}